\documentclass[12pt]{article}
\usepackage{graphicx}

\begin{document}

 \title{Influence of Phase Diffuser Dynamics on Scintillations of Laser Radiation in Earth Atmosphere:
Long-Distance Propagation}
\author{G.P. Berman $^{1}$\footnote{Corresponding
author: gpb@lanl.gov} and A.A. Chumak$^{1,2}$ 
\\[3mm]$^1$
 Los Alamos National Laboratory, Theoretical Division,\\ Los Alamos,
NM 87545
\\[5mm] $^2$  Institute of Physics of
the National Academy of Sciences\\ Pr. Nauki 46, Kiev-28, MSP 03028
Ukraine
\\[3mm]\rule{0cm}{1pt}}
\maketitle \markright{right_head}{LA-UR 09-01858}

\begin{abstract}

The effect of a random phase diffuser on fluctuations of laser light (scintillations) is studied. 
Not only spatial
but also temporal phase variations introduced by the phase diffuser are analyzed. The explicit 
dependence of the scintillation index on finite-time phase variations is obtained for long 
propagation paths. It is shown that for 
large amplitudes of phase fluctuations, a finite-time effect decreases the ability of phase 
diffuser to suppress the scintillations.


\end{abstract}







\section{Introduction}

Studies of laser beams propagating through turbulent atmospheres 
are important for many applications such as remote sensing, tracking, 
and long-distance optical communications. Howerver, fully coherent 
laser beams are very sensitive to fluctuations of the atmospheric 
refractive index. The initially coherent laser beam acquires some 
properties of Gaussian statistics in course of its propagation through 
the turbulence. As a result, the noise/signal ratio approaches unity for 
long-distance propagation. (See, for example, Refs.\cite{gra}-\cite{kor}).    
This unfavourable effect limits the performance of communication 
channels. To mitigate this negative effect the use of 
partially (spatially) coherent beams was proposed. The 
coherent laser beam can be transformed into a partially coherent beam 
by means of a phase diffuser placed near the exit aperture. This 
diffuser introduces an additional phase (randomly varying in space and time) 
to the wave front of the outgoing radiation. Statistical characteristics of the 
random phase determine the initial transverse coherence length 
of the beam.  It is shown in Refs. \cite{ban},\cite{ber} 
that a considerable decrease in the noise/signal ratio can occur under 
following conditions: (i) the ratio of the initial transverse coherence 
length, $\lambda_c$, to the beam radius, $r_0$, should be essentially 
smaller than unity; and (ii) the characteristic time of phase variations, $\tau_d$, 
should be much smaller than the integration time, $T$, of the detector. 
However, only limiting cases $(\tau_d/T)\rightarrow 0$ and 
$(\tau_d/T)\rightarrow \infty $ have been
 considered in the literature. (See, for example, Refs. \cite{ban},\cite{ber}
and Ref. \cite{ban2}, respectively). It is evident that the inequality $\tau_d<<T$ can 
be easily satisfied by choosing a detector with very long integration 
time. At the same time, this kind of the detector cannot distinguish different signals within 
the interval $T$. This means that the resolution of the receiving system might 
become too low for the case of large $T$. On the other hand, there is 
a technical restriction on
phase diffusers: up to now their characteristic times, $\tau_d$, are not smaller 
than $10^{-7} s$. Besides that, in some specific cases (see, for example, Ref. \cite{ber3}), the spectral 
broadening of laser radiation due to the phase diffuser ($\Delta \omega \sim \tau_d^{-1}$) 
may become unacceptably high. 

The factors mentioned above impose serious restrictions on the physical characteristics of 
phase diffusers which could be potentially useful for suppressing the intensity fluctuations. 
An adequate choice of diffusers may be facilitated if we know in detail the effect 
of finite-time phase variation, introduced by them, on the photon statistics. 
In this case, it is possible 
to control the performance of communication systems. In what follows, we will 
obtain theoretically the dependence of scintillation 
index on $\tau_d/T$ without any restrictions on the value of this ratio.
This is the main purpose of our paper. Further analysis is based on the 
formalism developed in Ref. \cite{ber} and modified here to understand the case of finite-time dynamics 
of the phase diffuser. 

\section{The method of photon distribution function in the problem of scintillations.}

The detectors of the absorbed type do not sense the instantaneous intensity of electromagnetic 
waves $I(t)$. They sense the intensity averaged over some finite interval $T$ i.e.
\begin{equation}\label{one}
\bar{I}(t)=\frac 1T\int_{t-T/2}^{t+T/2}I(t^\prime)dt^\prime .
\end{equation}
Usually, the averaging time $T$ (the integration time of the detector) is much smaller 
than the characteristic time of the turbulence variation, 
$\tau_A$, ($T<<\tau_A\sim 10^{-2}-10^{-3}s$). Therefore, the 
average value of the intensity can be obtained by further averaging of Eq. \ref{one} 
over many measurements corresponding various realizations of the refractive-index configurations. 

The scintillation index determining the mean-square fluctuations of the 
intensity is defined by 
\begin{equation}\label{two} 
 \sigma^2=\bigg [\frac 1{T^2}\int_{t-T/2}^{t+T/2}\int_{t-T/2}^{t+T/2}
dt^{\prime}dt^{\prime \prime}\big <:I(t^\prime)I(t^{\prime\prime}):\big >-\big <\bar{I}\big >^2\bigg ]\bigg /\big <\bar{I}\big >^2=
\frac{\big< :\bar I(t) ^2:\big>}{\big<\bar I \big>^2}-1,
\end{equation}
where the symbol $\{:...:\}$ indicates the normal ordering of the creation and annihilation operators 
which determine the intensity, $I(t)$. (See more details in Refs. \cite{ber},\cite{man}). The
brackets $<...>$ indicate quantum-mechanical and atmospheric averagings.

The intensity $I$ depends not only on $t$, but also on the spatial variable ${\bf r}$. 
Therefore, the detected intensity is the intensity $I({\bf r},t)$ averaged not only over 
$t$ as in Eq. \ref{one}, but also over the detector aperture. For simplicity, we will restrict 
ourselves to calculations of the intensity correlations for coinciding spatial points that 
correspond to "small" detector aperture. This simplification is 
quite reasonable for a long-distance propagation path of the beam. 

 In the case of quasimonochromatic light, we can choose $I({\bf r},t)$ in the form
\begin{equation}\label{three} 
 I({\bf r},t)=\frac 1V\sum_{\bf q,k}e^{-i{\bf kr}}b^+_{{\bf q}+{\bf k}/2}(t)
b_{{\bf q}-{\bf k}/2}(t),
 \end{equation}
where $b^+_{{\bf q}}(t)$ and $b_{{\bf q}}(t)$ are the creation and annihilation operators of 
photons with momentum ${\hbar \bf q}$.  They are given in the Heisenberg representation.
$V=L_xL_yL_z$ is the volume of the system.

It follows from Eqs. \ref{two},\ref{three} that $\sigma^2$ can be obtained if one knows the 
average
 \begin{equation}\label{four} 
 \big< :I({\bf r},t) I({\bf r},t^\prime):\big>=\frac 1{V^2}\sum_{\bf q,k,q^\prime ,k^\prime}e^{-i{\bf (k+k^\prime )r}}
 \big < b^+_{{\bf q}+\frac {\bf k}2}(t)b^+_{{\bf q^\prime}+\frac {\bf k^\prime}2}(t^\prime) 
b_{{\bf q^\prime}-\frac {\bf k^\prime}2}(t^\prime)b_{{\bf q}-\frac{\bf k}2}(t)\big>.
\end{equation}  
It is a complex problem to obtain this value for arbitrary turbulence strengths and propagation distances. 
Nevertheless, the following qualitative reasoning can help to do this in the case of strong  
turbulence. We have mentioned that the laser light acquires the properties of Gaussian 
statistics in the course of its propagation through the turbulent atmosphere. As a result, in the 
limit of infinitely long propagation path, $z$, only ``diagonal" terms, i.e. terms 
with (i) ${\bf k=k^\prime =0}$ or (ii) ${\bf q+k}/2={\bf q^\prime -k}^\prime /2$, 
${\bf q-k}/2={\bf q^\prime +k^\prime }/2$ contribute to the right part of 
Eq. \ref{four}. For large but still finite $z$, there exist small ranges of ${\bf k},{\bf k^\prime}\not=0$  
in case (i) and ${\bf q+k}/2\not={\bf q^\prime -k}^\prime /2$, 
${\bf q-k}/2\not={\bf q^\prime +k^\prime }/2$ in case (ii) contributing into the 
sum in Eq. \ref{four}. The presence of the mentioned regions is due to the two possible ways of 
correlating of four different waves (see Ref. \cite{das}) which enter the right hand side of 
Eq. \ref{four}. As explained in Ref. \cite{gor}, the characteristic sizes of regions (i) and (ii)
depend on the atmospheric broadening of beam radii as $(\Delta R_b)^{-1}$, 
thus decreasing with  
increasing $z$. In the case of long-distance propagation, $(\Delta R)^{-1}_b$ is much smaller 
than the component of photon wave-vectors perpendicular to the $z$ axis. The last quantity grows 
with $z$ as $z^{1/2}$. (See Ref. \cite{ber}). For this reason,  the overlapping 
of regions (i) and (ii) can be neglected. In this case Eq. \ref{four} can be rewritten in the convenient 
form: 
\begin{equation}\label{five} 
\big< :I({\bf r},t) I({\bf r},t^\prime):\big>=\frac 1{V^2}\sum_{\bf |k|,|k|^\prime <k_0}\sum_{\bf q,q^\prime}
\big <:e^{-i{\bf kr}}b^+_{{\bf q}+\frac {\bf k}2}(t)b_{{\bf q}-\frac {\bf k}2}(t)\times 
\end{equation}
\[e^{-i{\bf k^\prime r}}b^+_{{\bf q^\prime}+\frac{\bf k^\prime}2}
(t^\prime)b_{{\bf q^\prime}-\frac {\bf k^\prime}2}(t^\prime )+
e^{-i{\bf kr}}b^+_{{\bf q}+\frac {\bf k}2}(t)b_{{\bf q}-\frac {\bf k}2}(t^\prime )
e^{-i{\bf k^\prime r}}b^+_{{\bf q^\prime}+\frac{\bf k^\prime}2}
(t^\prime)b_{{\bf q^\prime}-\frac {\bf k^\prime}2}(t):\big >, \]
where the value $k_0$, confining summation over ${\bf k,k^\prime }$, is chosen to be greater than 
$R_b^{-1}$ but much smaller than the characteristic transverse wave vector of the photons; this is  
consistent with the above explanations. The two terms in the right-hand side correspond to the two regions of 
four-wave correlations.

The quantity 
\begin{equation}\label{six} 
f({\bf r},{\bf q},t)=\frac 1V\sum_{{\bf |k|} <k_0}
e^{-i{\bf kr}}b^+_{{\bf q}+\frac {\bf k}2}(t)b_{{\bf q}-\frac {\bf k}2}(t) 
\end{equation}
entering the right side of Eq. \ref{five} is the operator of photon density in phase space 
(the photon distribution function 
in ${\bf r,q}$ space). It was used in Refs. \cite{ber},\cite{gor} and \cite{qua} for 
the description of photon propagation in turbulent atmospheres. By analogy, we can 
define the two-time distribution function
\begin{equation}\label{seven} 
f({\bf r},{\bf q},t,t^\prime )=\frac 1V\sum_{{\bf |k|} <k_0}
e^{-i{\bf kr}}b^+_{{\bf q}+\frac {\bf k}2}(t)b_{{\bf q}-\frac {\bf k}2}(t^\prime ). 
\end{equation}
Then Eq. \ref{five} can be rewritten in terms of the distribution functions as
\begin{equation}\label{eight} 
\big< :I({\bf r},t) I({\bf r},t^\prime):\big>=\sum_{\bf q,q^\prime}
\big <:f({\bf r,q},t)f({\bf r,q^\prime},t^\prime )+f({\bf r,q},t,t^\prime )
f({\bf r,q^\prime},t^\prime ,t):\big> 
\end{equation}

Let us represent $f({\bf r},{\bf q},t,t^\prime )$ in the form $f({\bf r},{\bf q},t,t+\tau )$. 
We assume that $\tau \leq T\ll \tau_A$, as explained in the text after Eq.\ref{one}. 
In this case the Hamiltonian of photons in a turbulent atmosphere 
can be considered to be independent of time. As a result,  
both functions defined by Eqs. \ref{six} and \ref{seven} satisfy the same kinetic equation, i.e.   

\[(\partial _t+{\bf c_q}\partial _{\bf r}+{\bf F}\partial _{\bf q})f({\bf r},{\bf q},t)=0 \]
\begin{equation}\label{nine} 
(\partial _t+{\bf c_q}\partial _{\bf r}+{\bf F}\partial _{\bf q})f({\bf r},{\bf q},t,t+\tau )=0,
\end{equation}
where ${\bf c_q}=\partial \omega _{\bf q}/\partial {\bf q}$ is the photon velocity, ${\bf F}$ is 
a random force, caused by the turbulence. This force is equal to 
$\omega _0\partial n ({\bf r})/\partial {\bf r}$, where $\omega _0$ is the frequency of laser radiation.
$n({\bf r})$ is the refractive index of the atmosphere. 
The general solution of the equation for $f({\bf r},{\bf q},t,t+\tau )$ can be written in the form
\begin{equation}\label{ten} 
f({\bf r},{\bf q},t,t+\tau )=\frac 1V\sum_{{\bf |k|} <k_0}e^{-i{\bf kr}(t_0)}b^+_{{\bf q}(t_0)+
\frac {\bf k}2}(t_0)b_{{\bf q}(t_0)-\frac {\bf k}2}(t_0+\tau ),
\end{equation}
where 
\begin{equation}\label{eleven} 
{\bf r}(t_0)={\bf r}+\int_t^{t_0}{\bf\dot r}(t^\prime )dt^\prime
\end{equation}
\[{\bf q}(t_0)={\bf q}+\int_t^{t_0}{\bf\dot q}(t^\prime )dt^\prime .\]
The functions ${\bf r}(t^\prime )$ and ${\bf q}(t^\prime )$ obey the equations of motion
\begin{equation}\label{twelve}
{\bf \dot r}(t^\prime )={\bf c_q}(t^\prime ), {\bf \dot q}(t^\prime )={\bf F}({\bf r}(t^\prime ))
\end{equation}
with the boundary conditions ${\bf r}(t^\prime =t)={\bf r},{\bf q}(t^\prime =t)={\bf q}$. 
The instant $t_0$ is equal to $t-z/c$, where $c$ is the speed of light. $t_0$ is the 
time of the exit of photons from the source. This choice of $t_0$ makes it possible to neglect 
the influence of the turbulence on the initial values of operators $b^+(t_0),b(t_0+\tau )$ 
(their dependence on time is as in vacuum). 

The term for $f({\bf r},{\bf q},t)$ can be obtained from Eq. \ref{twelve} by putting $\tau =0$. 
Substituting both distribution functions into Eq. {\ref{eight}}, we obtain
\begin{equation}\label{thirteen}
\big< :I({\bf r},t) I({\bf r},t^\prime):\big>=\frac 1{V^2}\sum_{\bf |k|,|k|^\prime <k_0}
\sum_{\bf q,q^\prime}\big <:e^{-i{\bf kr_q}(t_0)-i{\bf k^\prime r_{q^\prime}}(t_0)}\times
\end{equation}
\[\big[b^+_{{\bf q}(t_0)+\frac {\bf k}2}(t_0)b_{{\bf q}(t_0)-\frac {\bf k}2}(t_0)
b^+_{{\bf q^\prime}(t_0)+\frac {\bf k^\prime}2}(t_0+\tau )b_{{\bf q^\prime}(t_0)-
\frac {\bf k^\prime}2}(t_0+\tau)+\]
\[b^+_{{\bf q}(t_0)+\frac {\bf k}2}(t_0)b_{{\bf q}(t_0)-\frac {\bf k}2}(t_0+\tau )
b^+_{{\bf q^\prime}(t_0)+\frac {\bf k^\prime}2}(t_0+\tau )b_{{\bf q^\prime}(t_0)-
\frac {\bf k^\prime}2}(t_0)\big]:\big>,\]
where ${\bf r_q(t_0)}$ and ${\bf r_{q^\prime}(t_0)}$ are solutions of Eqs. \ref{twelve} with the initial 
conditions ${\bf r}(t^\prime =t)={\bf r},{\bf q}(t^\prime =t)={\bf q}$ and ${\bf r}(t^\prime =t)=
{\bf r},{\bf q}(t^\prime=t)={\bf q^\prime}$, respectively.

\section{A phase diffuser with finite correlation time}

The operators on the right side of Eq. \ref{thirteen}  are related through matching conditions with the 
amplitudes of the exiting laser radiation (see Ref. \cite{ber}) by the relation
\begin{equation}\label{fourteen}
b_{{\bf q_\perp},q_0}=b(L_xL_y)^{-1/2}\int {\bf dr_\perp}e^{-i{\bf q_\perp r_\perp}}\Phi ({\bf r_\perp}),
\end{equation}
where $b$ is the operator of the laser field which is assumed to be a single-mode field and the subscript
($_\perp$) means perpendicular to the $z$-axis component. The function $\Phi$ describes the profile of the 
laser mode, which is assumed to be Gaussian-type function 
[$\Phi =\big(\frac 2\pi \big)^{1/2}r_0^{-1}e^{-r_\perp ^2/r_0^2}$]. 
$r_0$ desribes the initial radius of the beam. 

To account for the effect of the phase diffuser, a factor $e^{i\varphi ({\bf r}_\perp ,t_0)}$ or 
$e^{i\varphi ({\bf r}_\perp ,t_0+\tau)}$ should be inserted into the integrand of Eq. \ref{fourteen}. 
The quantity $\varphi ({\bf r}_\perp ,t)$ is the random phase introduced by the phase diffuser. A similar 
consideration is applicable to each of four photon operators entering both terms in square brackets of
Eq. \ref{thirteen}. It can be easily seen that the factor 
\begin{equation}\label{fifteen}
\Upsilon =e^{i[\varphi ({\bf r},t_0)-\varphi ({\bf r^\prime},t_0)+\varphi ({\bf r_1},t_0+\tau )-
\varphi ({\bf r_1^\prime },t_0+\tau )]},
\end{equation}
describing the effect of phase screen on the beam, enters implicitly the integrand of Eq. \ref{thirteen} 
(the indices $_\perp$ are omitted here for the sake of brevity). There are 
integrations over variables ${\bf r,r^\prime ,r_1,r_1^\prime}$ as shown in Eq. \ref{fourteen}. 
Furthermore, the brackets $<...>$, which indicate averaging over different realizations of the atmosperic 
inhomogeneities, also indicate averaging over different states of the phase diffuser. As long as both types
of averaging do not correlate, the factor (\ref{fifteen}) entering Eq. \ref{thirteen} must be averaged 
over different instants, $t_0$. To begin with, let us consider the simplest case of two phase 
correlations   
\begin{equation}\label{sixteen}
\big <e^{i[\varphi ({\bf r},t_0)-\varphi ({\bf r^\prime },t_0)]}\big >.
\end{equation}
It is evident that in the case $\langle|\varphi|\rangle>>1$, as shown schematically in Fig. 1, the factor
(\ref{sixteen}) is sizable if only points ${\bf r}$ and ${\bf r^\prime }$ are close to one another. 
\begin{figure}[t]
\centering
\includegraphics {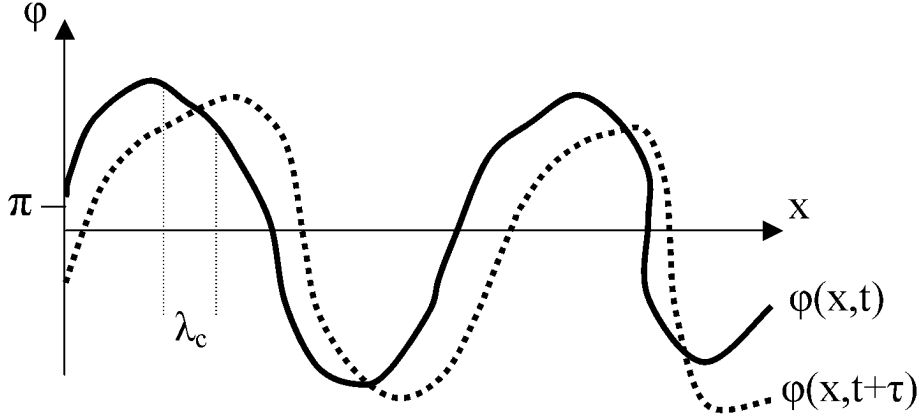}
\caption{Schematics of random phase variations along the $x$ direction. Two curves correspond to different 
instants $t$ and $t+\tau $.}
\end{figure}
Therefore,
the term given by Eq. \ref{sixteen} can be replaced by 
   \begin{equation}\label{seventeen}
\big <e^{i\frac {\partial \varphi ({\bf r},t_0)}{\partial {\bf r}}({\bf r-r^\prime })}\big >=
 e^{-\lambda _c^{-2}({\bf r-r^\prime })^2},
\end{equation} 
where $\frac {\partial \varphi ({\bf r},t_0)}{\partial {\bf r}}$ is considered to be a Gaussian random 
variable with the mean-square values given by 
$\langle [\frac {\partial \varphi ({\bf r},t_0)}{\partial x}]^2\rangle = 
\langle [\frac {\partial \varphi ({\bf r},t_0)}{\partial y}]^2\rangle =2\lambda _c^{-2}$, where $\lambda _c$ 
is the correlation length of phase fluctuations. (See Fig.1). As we see, in this case the effect of phase 
fluctuations can be described by the Schell model \cite{ban}-\cite{ban2},\cite{gbu}-\cite{shi}.  

A somewhat more complex situation is for the average value of $\Upsilon $ given by Eq. \ref{fifteen}. 
There is an effective phase correlation not only in the case of coincident times, but also for 
differing times. 
For $\lambda _c<<r_0$, two different sets of coordinates contribute considerably to phase correlations. 
This can be described mathematically as
\begin{equation}\label{eighteen}
\big <\Upsilon \big >=\big <e^{i[\varphi ({\bf r},t_0)-\varphi ({\bf r^\prime},t_0)+\varphi ({\bf r_1},t_0+\tau )-
\varphi ({\bf r_1^\prime },t_0+\tau )]}\big >\approx \big <e^{i[\varphi ({\bf r},t_0)-\varphi 
({\bf r^\prime},t_0)]}\big >\times 
\end{equation}
\[\big <e^{i[\varphi ({\bf r_1},t_0+\tau )-\varphi ({\bf r_1^\prime },t_0+\tau )]}\big >
+ \big <e^{i[\varphi ({\bf r},t_0)-\varphi 
({\bf r^\prime _1},t_0+\tau )]}\big > \big <e^{i[\varphi ({\bf r_1},t_0+\tau )-\varphi 
({\bf r^\prime },t_0)]}\big >.\]
Repeating the arguments leading to Eq. \ref{seventeen}, we represent the difference in the 
last term $\varphi ({\bf r},t_0)-\varphi ({\bf r^\prime _1},t_0+\tau )$ as
\begin{equation}\label{nineteen}
\frac {\partial \varphi ({\bf r},t_0)}{\partial {\bf r }}({\bf r-r_1^\prime })-
\frac {\partial \varphi ({\bf r},t_0)}{\partial t_0}\tau .
\end{equation}
Then, considering the random functions $\frac {\partial \varphi ({\bf r},t)}{\partial {\bf r }}$ and
$\frac {\partial \varphi ({\bf r},t)}{\partial t}$ as independent Gaussian variables, we obtain a
simple expression for $\big <\Upsilon \big >$. It is given by   
\begin{equation}\label{twenty}
\big <\Upsilon \big >=e^{-\lambda_c^{-2}[({\bf r-r^\prime})^2+({\bf r_1-r^\prime}_1)^2]}+
e^{-\lambda _c^{-2}[({\bf r-r^\prime}_1)^2+({\bf r^\prime -r_1})^2]-2\nu^2\tau ^2},
\end{equation}
where $\langle [\frac {\partial \varphi ({\bf r},t)}{\partial t}]^2\rangle =2\nu^2$.

As we see, the effect of the phase screen can be described by two parameters, $\lambda _c$ and $\nu $, 
which characterize the spatial and temporal coherence of the laser beam. In the 
limiting case,   
$\nu \rightarrow \infty$, the second term in Eq. \ref{twenty} vanishes and the problem is reduced 
to the case considered in Refs. \cite{ban},\cite{ber}. In the opposite case, $\nu \rightarrow 0 $,
both terms in Eq. \ref{twenty} are important. This is shown in Ref. \cite{ban2}. In what follows,
we will see that these two limiting cases have physical interpretations where
where $\nu T>>1$ (slow detector) 
and $\nu T<<1$ (fast detector), respectively.

There is a specific realization of the diffuser in which a random phase distribution moves across 
the beam. (This situation can be modeled by a rotating transparent disk with large diameter and
varying thickness.) The phase depends here on the only variable ${\bf r-v}t$, i.e. 
\begin{equation}\label{twon}
\varphi ({\bf r},t)\equiv \varphi ({\bf r-v}t),
\end{equation}
where ${\bf v}$ is the velocity of the drift. Then we have 
\begin{equation}\label{twtw}
\big <\Upsilon \big >\approx 
e^{-\lambda_c^{-2}[({\bf r-r^\prime})^2+
({\bf r_1-r^\prime}_1)^2]}+e^{-\lambda _c^{-2}[({\bf r-r^\prime_1+v}\tau)^2+({\bf r^\prime -r_1+v}\tau)^2]}.
\end{equation}
Comparing Eqs. \ref{twenty} and \ref{twtw}, we see that the quantity, $v/\lambda_c $, stands for the 
characteristic parameter describing the
efficiency of the phase diffuser. The criterion of ``slow" detector requires
$(vT/\lambda_c)>>1$. Qualitatively, the two scenarios of phase variations, given by 
Eqs. \ref{twenty} and \ref{twtw}, affect in a similar way
 the intensity fluctuations. In what follows, we consider the first of them as the simplest one. 
(This is because the spatial and temporal variables in $\big <\Upsilon \big >$, given by Eq. \ref{twenty},
 are separable.) 
\begin{figure}[ht]
\centering
\includegraphics{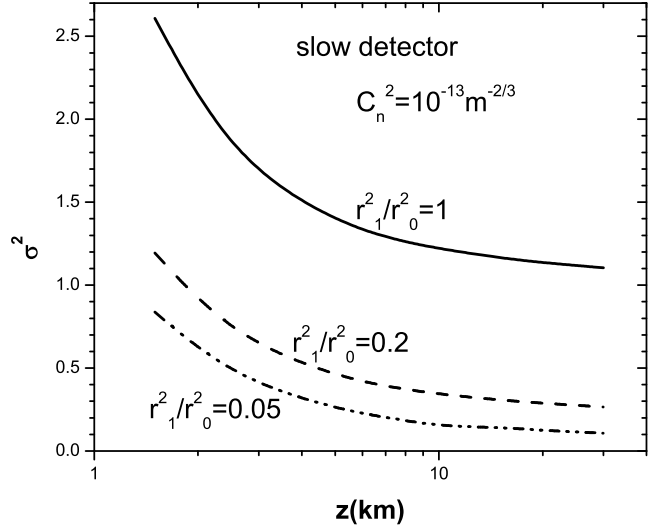}
\caption{The scintillation index $\sigma ^2$ {\it vs} propagation distance $z$ 
in the case of ``slow'' detector: $\nu T\rightarrow \infty $. The parameter
$r_1^2/r_0^2$ indicates different initial coherence length. In the absence of phase diffuser 
$(r_1/r_0)^2=1$ (solid line). $C^2_n$ is the conventional parameter describing a strength of the atmospheric 
turbulence.}
\end{figure}

Substituting the expressions for operators given by Eq. \ref{fourteen} with account for the phase
factors $e^{\pm i\varphi ({\bf r}_\perp ,t_0)}$ and averaging over time as shown in Eq. \ref{one}, 
we obtain 
\begin{equation}\label{twth}
\big <:{\bar I(t)}^2:\big >=\bigg (\frac {2\pi r_1^2}{L_xL_yV}\bigg )^2\big <b^+b^+bb\big >\Psi (\nu T)
\sum_{\bf |k|,|k|^\prime <k_0}
\sum_{\bf q,q^\prime}\bigg <e^{-i{\bf kr_q}(t_0)-i{\bf k^\prime r_{q^\prime}}(t_0)}\times
\end{equation}
\[\bigg[e^{(k^2+k^{\prime 2})r_0^2/8-(q_{t_0}^2+q_{t_0}^{\prime 2})r_1^2/2}+e^{(K_{t_0}^2+K_{t_0}^{\prime 2})
r_0^2/8-(Q_{t_0}^2+Q_{t_0}^{\prime 2})r_1^2/2}\bigg]\bigg >,\]
where the notation $<...>$ after sums indicates averaging over different realizations of the atmospheric 
refractive index. The parameter $r_1^2= r_0^2(1+2r_0^2\lambda _c^{-2})^{-1}$ describes the initial 
coherence length modified by the phase diffuser. Other notations are defined by following relations
\[\Psi (\nu T)\equiv  1+\frac {\Gamma (1/2)-\Gamma (1/2,2\nu^2T^2 )}{{\sqrt 2}\nu T}-
\frac {1-e^{-2\nu ^2T^2}}{2\nu^2T^2}\]
\[{\bf K}_{t_0}={\bf q}_{t_0}-{\bf q}_{t_0}^\prime +
\frac {\bf k+k^\prime}2,  {\bf K^\prime }_{t_0}={\bf q_{t_0}^{\prime }-q_{t_0}}+\frac {\bf k+k^\prime}2, \]
\[{\bf Q}_{t_0}=\frac {{\bf q}_{t_0}+{\bf q}_{t_0}^\prime }2+\frac {\bf k-k^\prime}4, 
{\bf Q}_{t_0}=\frac {{\bf q}_{t_0}+{\bf q}_{t_0}^\prime }2-\frac {\bf k-k^\prime}4,\]
Further calculations follow the scheme described in Ref \cite{ber}. Fig. 2 illustrates the effect of 
the phase 
diffuser on scintillations in the limit of a ``slow" detector ($\nu T\rightarrow \infty$). 
We can see a considerable decrease in $\sigma ^2$ caused by the phase diffuser.  
At the same time, the effect of the phase screen on $\sigma ^2$ becomes weaker for finite values of $\nu T$. 
Moreover, comparing the two upper curves in Fig. 3, we see the opposite effect: slow phase variations 
($\nu T =1$) result in increased  scintillations. There is a simple 
explanation for this phenomenon: the noise generated by the turbulence is complemented by the noise arising 
from the random phase screen. The integration time of the detector, $T$, is not sufficiently large for 
averaging phase variations generated by the diffuser.

The function, $\Psi (\nu T)$, has a very simple form in the two limits:
(i)$\Psi (\nu T)\approx 1+\frac {\sqrt \pi}{{\sqrt 2}\nu T}$, when $\nu T >>1$
; and (ii)  $\Psi (\nu T)\approx 2$, when $\nu T<<1$.
Then, in case (i) and for small values of the initial 
coherence [$(r_1^2/r_0^2)<<1$], the asymptotic term for the scintillation index ($z\rightarrow\infty$) is 
given by 
 \begin{equation}\label{twfo}
\sigma^2\approx \frac {r_1^2}{r_0^2}+\frac {\sqrt \pi}{\sqrt 2\nu T}. 
\end{equation}
The right-hand side of Eq. \ref{twfo} differs from analogous one in Ref. \cite{ber} by the value 
$\frac {\sqrt \pi}{{\sqrt 2}\nu T}$ that is much less than unity but, nevertheless, can be comparable 
or even greater than $(r_1/r_0)^2$.

In case (ii), the asymptotic value of $\sigma^2$ is close to unity, coinciding with the 
results of Refs. \cite{kor} and \cite{ban2}. This agrees with well known behavior of the scintillation 
index to approach unity for any source distribution, provided the response time 
of the recording instrument is short compared with the source coherence time. (See, for example, survey 
\cite{fan80}).
  
A similar tendency can be seen in both Figs. 3 and 4: the curves 
with the smallest $\nu T$, used for numerical calculations ($\nu T =1$), are close to the curves 
``without diffuser" in spite of the small initial coherence length [$(r_1/r_0)^2= 0.2; 0.05$]. 
It can also be seen that all curves approach their asymptotic values very slowly.  
\begin{figure}[ht]
\centering
\includegraphics{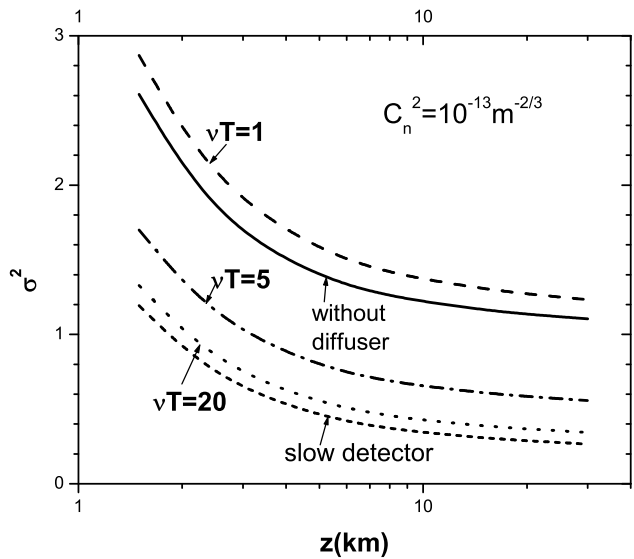}
\caption{Dependence of scintillation index on the distance z for different values of the parameter 
$\nu T$ describing diffuser dynamics. The solid curve is calculated for $(r_1/r_0)^2=1$ (without diffuser). 
Other curves are for $(r_1/r_0)^2=0.2$.}
\end{figure}
\begin{figure}[ht]
\centering
\includegraphics{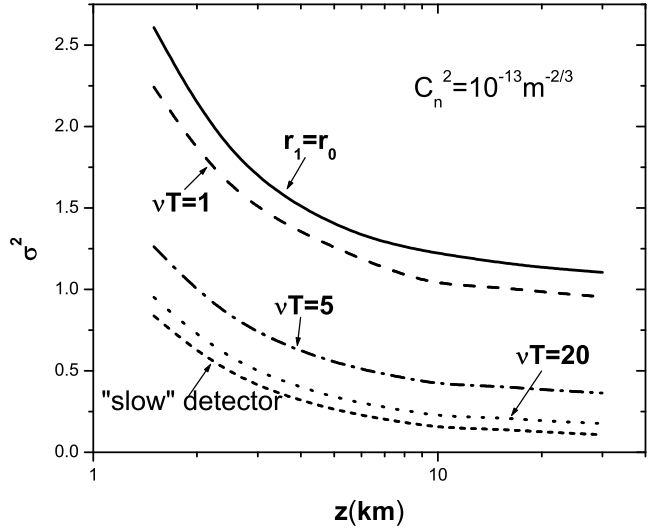}
\caption{The same as in Fig. 3 but for smaller initial coherence length: $(r_1/r_0)^2=0.05$.}
\end{figure}
\section{Discussion}
It follows from our analysis that the scintillation index is very sensitive to the diffuser 
parameters, $\lambda_c$ and $\nu$, for long propagation paths. On the other hand, the characteristics 
of the irradience such as beam radius, $R$, 
and angle-of-arrival spread, $\Delta \theta$, do not depend on the presence of the phase diffuser for large 
values of $z$. To see this, the following analysis is useful.

The beam radius expressed in terms of the distribution function is given by  
\begin{equation}\label{twfi}
R^2(z)=\frac {\sum_{\bf q}\int d{\bf r_\perp}r^2_\perp \big <f({\bf r},z,{\bf q},t)\big >}
{\sum_{\bf q}\int d{\bf r_\perp} \big <f({\bf r_\perp },z,{\bf q},t)\big >}.
\end{equation}
Straightforward calculations using Eq. \ref{ten} with $\tau =0$ (see Ref. \cite{ber}) result in 
the following explicit form:
\begin{equation}\label{twsi}
R^2(z)=\frac {r_0^2}2+\frac {2z^2}{q_0^2r_1^2}+4Tz^3,
\end{equation}
where $T=0.558C_n^2l_0^{-1/3}$ and $l_0$ is the inner radius of turbulent eddies, which in our previous
calculations was assumed  to be equal $2\pi \times 10^{-3}$m. As we see, the third term 
does not depend on the diffuser parameters and it dominates when $z\rightarrow \infty $. 

A similar situation holds for the angle-of-arrival spread, $\Delta \theta$. (This physical quantity 
is of great importance for the performance of 
communication systems based on frequency encoded information \cite{ber3}.) It is defined by the 
distribution function as
\begin{equation}\label{twse}
(\Delta \theta ({\bf r_\perp},z))^2=\frac {\sum_{\bf q}q^2_\perp 
\big <f({\bf r_\perp},z,{\bf q},t)\big >}
{q_0^2\sum_{\bf q} \big <f({\bf r_\perp },z,{\bf q},t)\big >}.
\end{equation}
Simple calculations, which are very similar to those while obtaining $R^2$,  result in  
\begin{equation}\label{twei}
[\Delta \theta (r_\perp =0,z)]^2=\frac 2{r_1^2q_0^2}+12Tz-\frac {4z^2}{q_0^4R^2}(r_1^{-2}+3Tq_0^2z)^2.
\end{equation}
For long propagation paths, Eq. \ref{twei} reduces to $3Tz$, which like $R^2$ 
does not depend on the diffuser parameters. 

As we see, for large distances $z$, the quantities $R^2$ and $\Delta \theta$ do not depend on 
$\lambda _c$ and $\nu $. This contrasts with the case of the scintillation index. So pronounced 
differences can be explained by differences in the physical nature of these characteristics. It follows
from Eq. \ref{two} that the functional, $\sigma^2$, is  quadratic in the distribution function, $f$. 
Hence,
four-wave correlations determine the value of scintillation index. The main effect of a phase diffuser
on $\sigma^2$ is to destroy correlations between waves exited at different times. (See more explanations 
in Ref. \cite{qua}). This is achieved at  
sufficiently small parameters $\lambda _c$ and $\nu ^{-1}$. 

In contrast, $R^2$ and $\Delta \theta$ depend 
on two wave-correlations, both waves being given at the same instant. Therefore, the values of $R^2$ 
and $\Delta \theta$ do not depend on the rate of phase
variations [$\nu $ does not enter the factor 
 (\ref{seventeen}) describing the effect of phase diffuser]. 
Moreover, these quantities become independent of $\lambda _c$ at long propagation paths because 
light scattering on atmospheric inhomogeneities prevails in this case. 

The plots in Figs. 3 anf 4 show that the finite-time effect is quite sizable even for very ``slow" 
detectors ($\nu T =20$). Our paper makes it possible to estimate the actual utility of phase 
diffusers in several physical regimes.


\section{Conclusion}
We have analyzed the effects of a diffuser on scintillations for the case of large-amplitude phase 
fluctuations. This specific case is very convenient for theoretial analysis because only two parameters 
are required to describe the effects of the diffuser. Phase fluctuations may occur independently 
in space as well as in time. Also, our formalism can be applied for the physical situation
in which a spatially random phase distribution drifts across the beam. (See Eq. \ref{twtw}.) Our results 
show the importance of both parameters, $\lambda _c$ and $\nu$, on the ability of a phase diffuser to
suppress scintillations. 

\section{Acknowledgment}
This work was carried out under the auspices of the
National Nuclear Security Administration of the U.S. Department of
Energy at Los Alamos National Laboratory under Contract No.
DE-AC52-06NA25396. We thank ONR for supporting this research.

\newpage \parindent 0 cm \parskip=5mm





\end{document}